\documentclass[conference,hidelinks]{IEEEtran}
\usepackage{amsmath,graphicx}
\usepackage{url}
\usepackage{amsfonts}
\usepackage{amssymb}
\usepackage{amsbsy}
\usepackage{bm}
\usepackage{booktabs}
\usepackage{times}
\usepackage{color}
\usepackage[noadjust]{cite}
\usepackage{enumerate}
\usepackage{enumitem}
\usepackage{tikz}
\usepackage{graphicx}
\usepackage{amsthm}
\usepackage{dsfont}
\usepackage{algpseudocode}
\usepackage{algorithm}
\usepackage{tabularx}
\usepackage{placeins}
\usepackage[para]{threeparttable}
\usepackage{multirow}
\usepackage{multicol}
\usepackage{setspace}
\usetikzlibrary{arrows}
\usepackage[binary-units]{siunitx}
\usepackage{paralist}

\usepackage[noadjust]{cite}

\usepackage{footnote}
\makesavenoteenv{tabular}
\makesavenoteenv{table}
\usepackage{soul}

\usepackage{pgfplots}
\pgfplotsset{compat=1.14}

\newcolumntype{C}{>{\centering\arraybackslash}X}
\newcolumntype{R}{>{\raggedleft\arraybackslash}X}
\newcolumntype{x}[1]{>{\centering\arraybackslash\hspace{0pt}}p{#1}}

\newcommand{\secref}[1]{{Sec.~\ref{#1}}}

\newcommand{\figref}[1]{{Fig.~\ref{#1}}}
\newcommand{\tabref}[1]{{Tbl.~\ref{#1}}}
\newcommand{\algoref}[1]{{Alg.~\ref{#1}}}

\title{Hydra: An Accelerator for Real-Time Edge-Aware Permeability Filtering in 65nm CMOS\vspace{-2mm}}
\author{
\IEEEauthorblockN{M. Eggimann$^\dag$, C. Gloor$^\dag$, F. Scheidegger$^\dag$, L. Cavigelli$^\dag$, M. Schaffner$^\dag$, A. Smolic$^\ddag$, L. Benini$^\dag$}
\IEEEauthorblockA{$^\dag$ETH Zurich, Integrated Systems Lab IIS, Zurich, Switzerland \hspace{0.5cm} $^\ddag$Trinity College, Dublin, Ireland
\vspace{-4mm}
}
}

\begin{document}
\bstctlcite{IEEEexample:BSTcontrol}
\maketitle
%
\begin{abstract}
Many modern video processing pipelines rely on edge-aware (EA) filtering methods. However, recent high-quality methods are challenging to run in real-time on embedded hardware due to their computational load. To this end, we propose an area-efficient and real-time capable hardware implementation of a high quality EA method. In particular, we focus on the recently proposed permeability filter (PF) that delivers promising quality and performance in the domains of HDR tone mapping, disparity and optical flow estimation. We present an efficient hardware accelerator that implements a  tiled variant of the PF with low on-chip memory requirements and a significantly reduced external memory bandwidth (6.4$\times$ w.r.t. the non-tiled PF). The design has been taped out in \SI{65}{\nano\metre} CMOS technology, is able to filter 720p grayscale video at \SI{24.8}{\hertz} and achieves a high compute density of \SI{6.7}{\giga FLOPS/\milli\metre\squared} (12$\times$ higher than embedded GPUs when scaled to the same technology node). The low area and bandwidth requirements make the accelerator highly suitable for integration into SoCs where silicon area budget is constrained and external memory is typically a heavily contended resource.

\end{abstract}

\vspace{-0.5mm}
\section{Introduction}
\label{sec:intro}
\vspace{-0.5mm}
Edge-aware (EA) filters are important building blocks used in many image-based applications like stylization, HDR tone mapping, detail editing and noise reduction \cite{Tomasi1998,Perona1990,Milanfar2013,Farbman2008,Fattal2009,He2013,Gastal2011DT,Gastal2015,Aubry2014,Paris2011,cigla2013,Aydin2014}. However, several high-quality methods such as the \emph{weighted least squares} (WLS) filter \cite{Farbman2008} are computationally demanding and hence unsuitable for real-time applications on resource constrained devices.
We focus on a recently proposed method termed \emph{permeability filter} (PF) \cite{cigla2013} that can be used to approximate image-based regularization problems such as HDR tone mapping \cite{Aydin2014}, disparity \cite{cigla2013}, and optical flow estimation \cite{Schaffner2017}.
The PF has been designed to converge to results similar to the high-quality WLS filter, but with significantly lower computational effort \cite{Schaffner2017} -- which renders the PF an ideal candidate for high-quality filtering in real-time.

In this work, we present a hardware accelerator for the PF that can be used as an area-efficient co-processor in systems-on-chip (SoCs) tailored towards video processing. In particular, we contribute the following:
\begin{compactitem}
\item We propose a tiled variant of the PF (TPF) with low on-chip memory requirements and a 6.4$\times$ lower off-chip memory bandwidth than the non-tiled PF.
\item We devise an efficient hardware architecture for the TPF that employs loop pipelining and an optimized memory interleaving scheme. Our design maximizes floating-point (FP) unit utilization and eliminates memory contentions caused by frequent tile transpositions required by the TPF.
\item We implement custom FP arithmetic on our accelerator to accurately filter feature maps involving HDR and coordinate data (e.g., sparse optical flow \cite{Lang2012,Schaffner2017}). 
\item Our design is the first custom hardware implementation taped-out in \SI{65}{\nano\metre} CMOS technology, and provides a high compute density of \SI{6.7}{\giga FLOPS/\milli\metre\squared}. When scaled to \SI{16}{\nano\metre} technology, this is around 12$\times$ denser than in recent embedded GPUs. When applying 4 internal PF iterations the chip processes 720p monochromatic video at \SI{24.8}{\hertz} with a measured power of \SI{445}{\milli\watt}.
\end{compactitem}
%
\section{Related Work}
The PF approximates the high-quality WLS filter \cite{Farbman2008} with a low computational effort \cite{cigla2013,Aydin2014,Schaffner2017}. Furthermore, the PF features good halo reduction and information spreading capabilities that are important for HDR tone-mapping, regularization methods, sparse-to-dense conversions, disparity and optical flow estimation. Other EA filters such as variants of the \emph{bilateral filter} (BF) \cite{Porikli2008} and the \emph{guided filter} (GF) \cite{He2013} are computationally less involved as the PF, but do not achieve the same level of quality and are hence used for different applications. We compare our chip with ASIC implementations of the BF \cite{Tseng2011} and GF \cite{Kao2014} accelerators in \secref{sec:results}. 
%
\section{Permeability Filter and Tiling}
\label{sec:perm_filter}
Similar to other EA filtering methods such as the GF and the \emph{domain-transform}, the PF uses a \emph{guiding image} $\mathbf{I}$ that controls the EA filtering behavior. The filtered data channels $\mathbf{A}$ may differ from the input image (e.g., in certain applications, $\mathbf{A}$ may hold other features like sparse optical flow vectors or disparity data). In a first step, the PF algorithm extracts \emph{pairwise permeabilities} $\pi^X_{pq}$ and $\pi^Y_{pq}$ from the guiding image $\mathbf{I}$ \cite{Schaffner2017}. Permeabilities measure the similarity between pixels at index $p$ and $q$ in the horizontal and vertical direction, and define the row-stochastic matrices $H^X_{pq}$ and $H^Y_{pq}$ holding the filtering coefficients for the horizontal and vertical filter passes. The PF is defined as a 1D operation over a \emph{single row/column} in the image as follows:
\vspace{-0.5mm}
\begin{equation} \label{eq:baseline_perm_filter}
    J_p^{(k+1)}=\sum_{q=1}^n H_{pq}J_q^{(k)}+\lambda H_{pp}(A_p - J_p^{(k)}),
\end{equation} 
%
where $\mathbf{A}$ holds the data channel to be filtered, $\mathbf{J}^{(k)}$ denotes the intermediate filtering result after $k$ iterations ($\mathbf{J}^{(0)}=\mathbf{A}$), $\lambda$ is a bias parameter towards the original data to reduce halo artifacts and $n$ denotes the length of the current row/column to be filtered. 
To generalize the PF into two dimensions, we iteratively apply \eqref{eq:baseline_perm_filter} on each row (called \emph{X-Pass}) and column (called \emph{Y-Pass}). Typical applications considered in this work (HDR tone-mapping \cite{Aydin2014}, filtering of optical flow data \cite{Schaffner2017}) apply $K=4$ \emph{XY-passes} to the entire frame.

Reformulating Equation \eqref{eq:baseline_perm_filter} enables an efficient 1D scanline evaluation \cite{cigla2013,Schaffner2017}. Each \emph{X-Pass} and \emph{Y-Pass} is decomposed into a forward and backward recursion. The forward recursion $F_{i}$ with normalization weight $\hat{F}_i$ is given by:
\begin{equation} \label{eq:forwardPass}
    F_{p} = \pi_{p-1}(F_{p-1} + J^{(k)}_{p-1}), \qquad \hat{F}_p = \pi_{p-1}(\hat{F}_{p-1} + 1.0),
\end{equation} 
and a backward recursion $B_{p}$ with normalization weight $\hat{B}_p$:
\begin{equation} \label{eq:backwardPass}
    B_{p-1} = \pi_{p-1}(B_p + J^{(k)}_p), \quad \hat{B}_{p-1} = \pi_{p-1}(\hat{B}_p + 1.0). 
\end{equation}
The $\bm{\pi}$ map in Equations~\eqref{eq:forwardPass} and \eqref{eq:backwardPass} is either $\bm{\pi}^X$ or $\bm{\pi}^Y$ depending on the filtering direction. The initial values of the recursions are set to zero. Finally, the resulting filter output can be computed on-the-fly during the backward recursion by
\begin{equation} \label{eq:combineFB}
    J_p^{(k+1)} = \frac{F_p + J^{(k)}_p + B_p + \lambda(A_p - J^{(k)}_{p})}{\hat{F}_p + 1.0 + \hat{B}_p}. 
\end{equation}
Hence, one PF iteration comprises a forward/backward recursion in x-direction, followed by a forward/backward recursion in y-direction, as illustrated in \figref{fig:tiling}a.
\subsection{Image Tiling}
\label{sec:imageTiling}
\begin{figure}[t]
    \centering
    \includegraphics[width=0.48\textwidth]{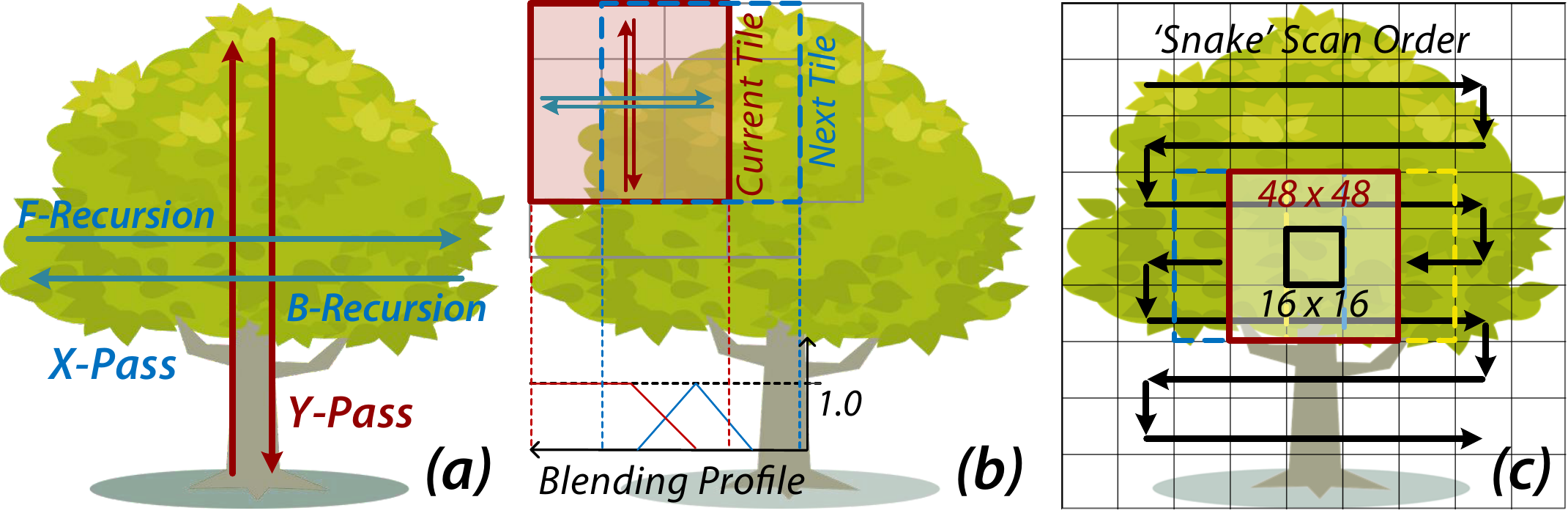}
    \caption{(a) The PF filters the frame by alternating between horizontal and vertical 1D filtering passes over the whole frame. (b) The TPF splits the frame into overlapping tiles that are filtered individually. The individual results are merged together using a linear weighting profile. (c) Row-major `snake' traversal scheme of the tiles.
    \vspace{-3mm}}
    \label{fig:tiling}
\end{figure}
\begin{figure}[t]
    \centering
    \includegraphics[width=0.49\textwidth]{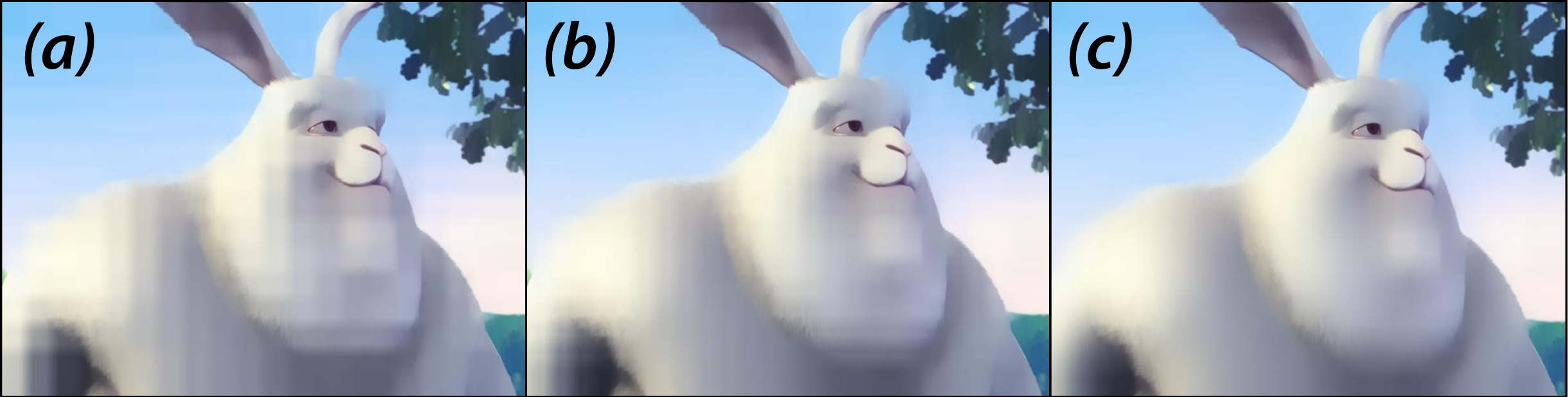}
    \caption{(a) Filter result with an overlap of $1/2$. (b) Filter result with an overlap of $3/5$. (c) Filter result with an overlap of $2/3$.
    \vspace{-3mm}}
    \label{fig:traversal_overlap_ratio}
\end{figure}
\begin{figure}[!b]
    \small
    \centering
    
    %
    
    \vspace{-3mm}      
    \includegraphics[width=0.36\textwidth]{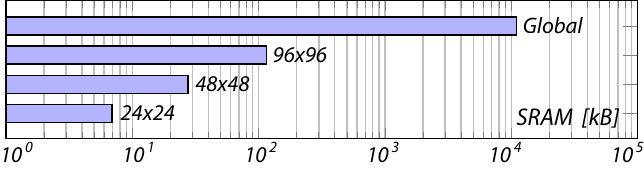}
    \vspace{-3mm}    
    \caption{Memory requirements for different tile sizes assuming 24 bit FP values (without temporary storage for intermediate results in the $F$, $B$ recursions).}
    \label{fig:tileSize}
    \vspace{-3mm}
\end{figure}

Since the PF alternates operating on rows and columns, the complete frame must be kept in the working memory. When applying the PF globally to one data channel $\mathbf{A}$ of a 720p frame ($h=720$ and $w=1280$) with a word width of $b=\SI{24}{\bit}$ (see \secref{sec:precision} for an evaluation) the required working memory amounts to $4\cdot w\cdot h\cdot b\approx\SI{88.5}{\mega\bit}$ to hold the two permeability maps, $\mathbf{A}$ and $\mathbf{J}^{(k)}$ (without intermediate storage for $\mathbf{F},\mathbf{\hat{F}}$). Since we are considering a co-processor scenario, such a large memory is unfeasible to be implemented on-chip, and hence requires off-chip memory. However, this results in a large off-chip memory bandwidth of $11\cdot K\cdot 2\cdot w\cdot h\cdot b\cdot\theta\approx\SI{6.09}{\giga\byte/\second}$ for $K=4$ filter iterations and a throughput of $\theta=\SI{25}{fps}$, which is not desirable. 
%
%
To this end, we propose a localized version of the PF, which operates on local square tiles as illustrated in \figref{fig:tiling}b. To reduce tiling artifacts, we employ linear blending of neighboring tiles. By evaluating different overlap configurations, we found that overlapping tiles by 2/3 provides the best visual quality (see \figref{fig:traversal_overlap_ratio}). With that configuration, nine different tiles contribute to result pixels. To minimize the memory requirements, we employ a `snake' scan order (illustrated in \figref{fig:tiling}c) to be able to reuse intermediate results for blending. \figref{fig:tileSize} shows on-chip SRAM requirements caused by different tile sizes.

A larger tile size is desirable to better approximate the global filter behavior. The following considerations restrict the choice of the tile size: tiles should overlap by 2/3 edge lengths, the length must be divisible by three, and computing the linear weights for the final merging step is simplified when the length is divisible by a power of two. This results in a preferred tile size of $3\cdot2^l\times3\cdot 2^l$. We choose a tile size of $48 \times 48$ pixels, of which $32 \times 48$ pixels overlap the neighbouring tiles on each side. Using this tiling approach, the PF can be reformulated to \algoref{alg:filter}, which can be implemented with only $4\cdot 48^2 \cdot b \approx\SI{27.6}{\kilo\byte}$ SRAM storage to hold one tile. Further, the external bandwidth, comprising the input data $\mathbf{A}$, $\bm{\pi}^X$, $\bm{\pi}^Y$, the filter output $\mathbf{J}^{(K)}$, and the partially blended tiles, reduces by 6.4$\times$ to only \SI{950}{\mega\byte/\second}.
\vspace{-3mm}
\begin{algorithm}[htb]
\setstretch{0.9}
\caption{Scanline Permeability Filter}
\begin{algorithmic}
\ForAll {Tiles}
    \For{ $1:(2\times k$) }
        \ForAll {Rows}
            \State Initialize recursions.
            \State Evaluate forward recursion.     \Comment{Eq. \eqref{eq:forwardPass}}
            \State Evaluate backward recursion.    \Comment{Eq. \eqref{eq:backwardPass}}
            \State Compute filter output.          \Comment{Eq. \eqref{eq:combineFB}}
         \EndFor
        \State Transpose tile.
    \EndFor 
    \State Blend overlapping tiles (linear profile). \Comment{\figref{fig:tiling}b}.
\EndFor
\end{algorithmic}
\label{alg:filter}
\end{algorithm}
\vspace{-5mm}
\begin{figure*}
    \centering
    \includegraphics[width=0.94\textwidth]{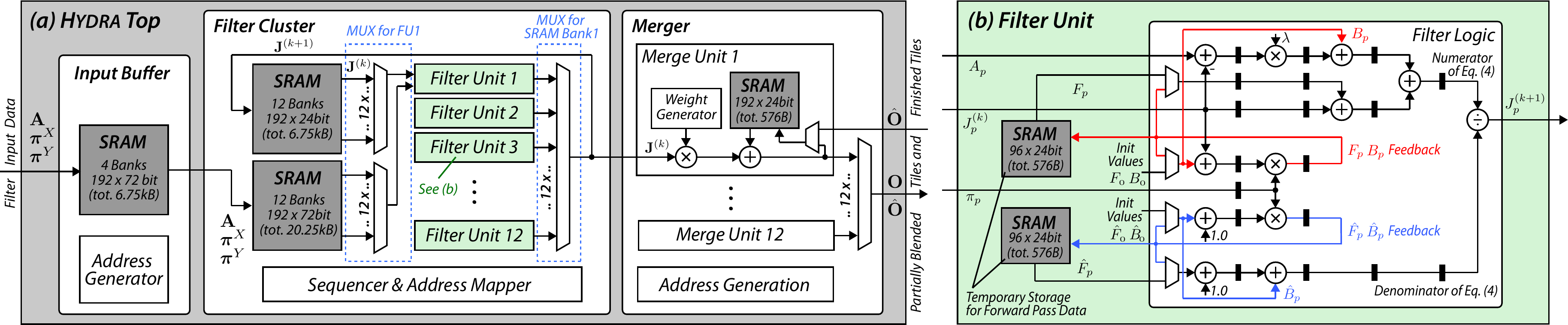}
    \vspace{-2mm}       
    \caption{(a) The \textsc{Hydra} chip contains an input buffer, a filter cluster with 12 FUs, and a merger unit. (b) The systolic FU datapath implements the forward/backward recursions $F_p$, $\hat{F}_p$, $B_p$ and $\hat{B}_p$, and employs pipeline interleaving to efficiently utilize the FP operators , as described in \secref{sec:filterCluster}.}
    \vspace{-3mm}
    \label{fig:toplevelArch}
\end{figure*}
%
\subsection{Numerical Precision}
\label{sec:precision} 
\begin{figure}[!b]
    \small
    \centering
    
    
    
    \includegraphics[width=0.4\textwidth]{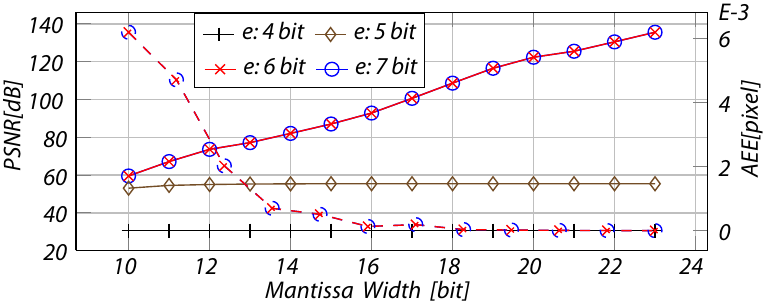}
    \vspace{-3mm}   
    \caption{PSNR (solid) and AEE (dashed) of the filter results for various floating point formats (exponent, mantissa width) compared to double precision.}
    \label{fig:psnr}
\end{figure}

One use-case of the PF algorithm is to regularize sparse feature maps (e.g., optical flow vectors) and convert them to a dense representation. This operation requires high precision and dynamic that is difficult to handle with fixed-point arithmetic. On the other hand, single precision FP with full IEEE-754 support (denormals, NaN, etc) is not needed for this application. \figref{fig:psnr} shows an evaluation of different FP formats for dense image data, as well as for the optical-flow estimation procedure \cite{Schaffner2017} that operates on sparse velocity data. Result quality is measured w.r.t. a double precision baseline with the PSNR measure in the case of dense data, and with the average endpoint error (AEE) measure for sparse flow data. Exponents with 5\,bit and below often lead to underflows for both data types and were hence not further considered. We chose to employ a 24\,bit FP format (FP24) with 6 exponent and 17 mantissa bits in order to align the format to byte boundaries (a byte aligned 16\,bit FP format would have led to unacceptable quality losses for both dense and sparse data). This leads to a negligible implementation loss below 2E-4 AEE for sparse flow data, and over 90\,dB PSNR for dense image data. 
\section{Architecture}
\label{sec:architecture}
\figref{fig:toplevelArch}a shows the proposed TPF architecture consisting of \emph{input buffer}, \emph{filter cluster} and \emph{merger}.
The \emph{tiles} of the current frame are streamed through the \emph{input buffer} into the \emph{filter cluster} that operates on one 48$\times$48 tile at a time. The \emph{input buffer} aggregates the input such that it can be bursted into the \emph{filter cluster} together with the last y-pass of the currently processed tile. I.e., the computation of the last y-pass is effectively overlapped with the data transfers that replace the now obsolete data in the two-port tile memories. During the last y-pass, the filter output is streamed into the \emph{merger} unit that fuses the overlapping tile areas and finally outputs the results. The architecture implements the \emph{tiled PF} with a fixed number of $K=4$ iterations, and is parameterized to process monochromatic images with 720p resolution in real-time. However, the same design can be scaled to higher resolutions and multiple channels by deploying more parallel filter units (FUs). By minimizing the bandwidth to external memory, we maximize energy efficiency and facilitate integration into a larger system (e.g., as a filter accelerator in a mobile SoC). 
\subsection{Filter Cluster}
\label{sec:filterCluster}
Due to the filter feedback loop, FP24 units should operate at single cycle latency to achieve high utilization (\secref{sec:interleaving}). Core frequencies up to 300 MHz achieve single cycle operation. Henceforth, we assume the limiting frequency of 300 MHz in the following throughput calculations.
The 1D PF requires a single FP division per pixel. One 720p frame is split into 3354 tiles. Each pixel in a single tile is processed 8 times (4 iterations with 1 horizontal and 1 vertical pass each). In order to achieve a throughput of 25\,fps, we need $25 \times 3354 \times 48^2 \times 8 \approx 1.55 \cdot 10^9$ FP divisions per seconds. Since the divisions are only performed during the backward recursion, we need at least $2 \times 1.55\;\mathrm{GFLOPS}/300\;\mathrm{MHz} \approx 10$ FP dividers to run in parallel. The proposed architecture hence contains 12 parallel FUs. As described in the forthcoming sections, we employ pipeline interleaving and an optimized SRAM interleaving pattern to achieve a high utilization rate of $99.9\%$ for FP multipliers and adders (49.9\% utilization rate for the FP dividers), that is required to achieve the targeted throughput without further datapath replication.

    \subsection{Pipeline Interleaving}
    \label{sec:interleaving}
    
    \begin{figure}[ht]
        \centering
         \vspace{-4mm}     
        \includegraphics[width=0.4\textwidth]{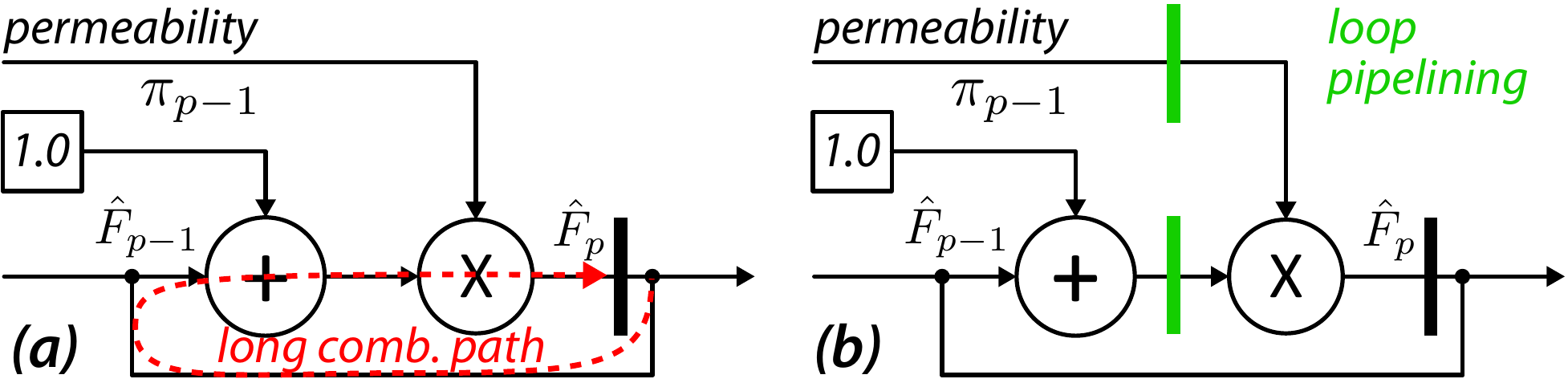}
        \vspace{-2mm}           
        \caption{Recursive part without (a) and with (b) pipeline register in the loop.}
        \label{fig:datapath_interleaving}
        \vspace{-2mm}
    \end{figure}
    \figref{fig:datapath_interleaving}a shows part of the datapath inside the FUs. Due to the long combinational path through the FP adder and multiplier, timing closure can not be achieved at the target frequency of \SI{300}{\mega\hertz}. The insertion of a pipeline register as in \figref{fig:datapath_interleaving}b improves the timing, but the feedback path from the multiplier back to the adder leads to a different functional behavior if the additional latency is not accounted for. To this end, a technique called \emph{pipeline interleaving} is used \cite{Kaeslin2014}. Instead of processing a single row or column of the tile at a time, each FU simultaneously processes two lines in an interleaved manner. As can be seen in \figref{fig:checkerboard_pixelPerm}a, the next pixel from an even row enters the FU in even cycles, and the next pixel from the neighboring odd row enters the FU in odd cycles. With the additional pipeline stage the propagation delay of the critical path in the FUs can be reduced to $3 \textnormal{ns}$.
    \subsection{Data Access Pattern}
    The proposed architecture provides simultaneous access to currently processed pixels in all operation modes and avoids filter pipeline stalls. To load twelve pixels in parallel, at least one memory block per filter block is required.
    Storing full rows of the tile in different memory blocks allows parallel access to the pixels during the horizontal pass but prevents simultaneous processing of the first pixels in the vertical pass since all the pixels of the first row reside in the same memory block. Instead, we employ an access pattern that subdivides the tile into squares of $2 \times 2$ pixels denoted as $s_{i,j}$. The rule $m(s_{i,j})=\mod\left(i+j,120\right)$ assigns squares to memory blocks $\in [0,...,11]$.
    The square size of $2 \time 2$ pixels is motivated due to the \emph{pipeline interleaving} and reduces the complexity of the address and memory index calculation units. The resulting checkerboard-like pattern is visualized in \figref{fig:checkerboard_pixelPerm}.
    \subsection{Cyclic Permutation of Pixel Locations}
    \begin{figure}[tb]
        \centering
        \includegraphics[height=0.20\textwidth]{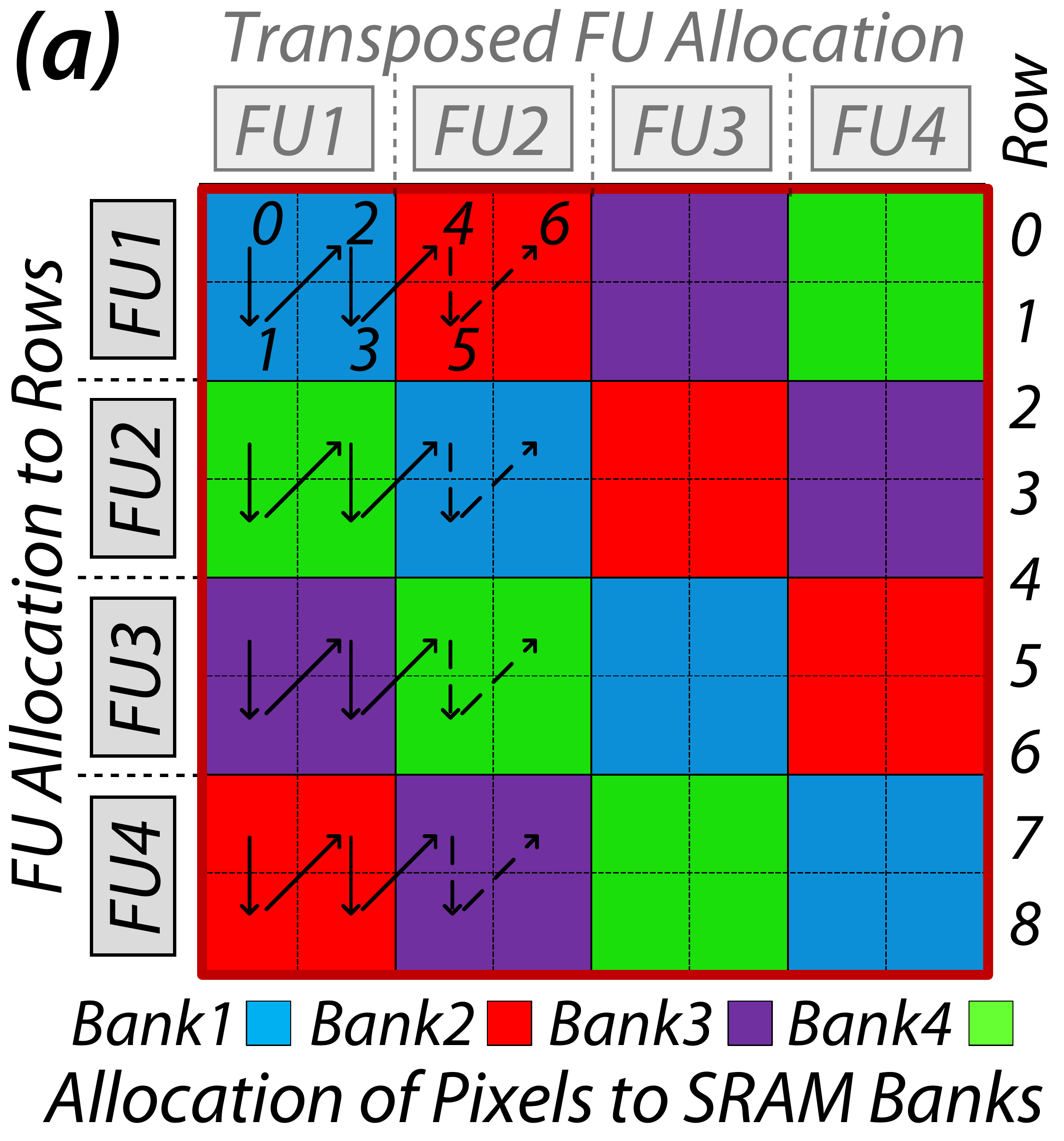}
        \hspace{2mm}
        \includegraphics[height=0.20\textwidth]{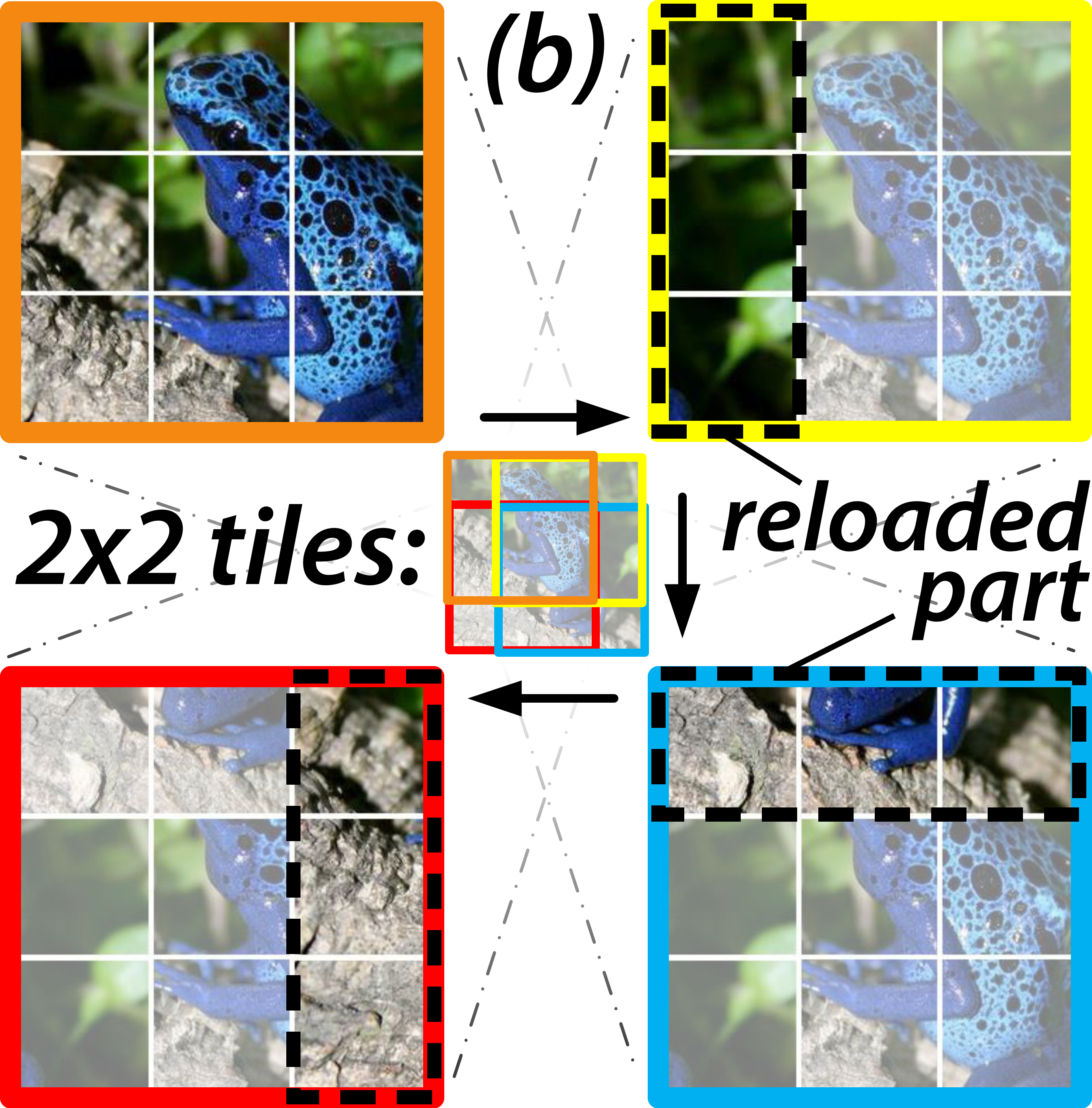}
        \vspace{-2mm}   
        \caption{(a) SRAM interleaving for stall-free tile transposition (shown for 4 FUs). (b)~Tile fragmentation and cyclic replacement scheme. Note that stepping from one tile to the next only requires to reload 1/3 of the tile.\vspace{-4mm}}
        \label{fig:checkerboard_pixelPerm}
    \end{figure}
    The proposed architecture tiles frames such that every two adjacent tiles overlap by 2/3, as explained in \secref{sec:imageTiling}. The tile size of $48 \times 48$ pixels implies -- with the exception of the first tile of a frame -- that $16 \times 48$ pixels (or $48 \times 16$ in the case of switching rows) need to be replaced. Reusing the remaining pixels reduces the input bandwidth. To maximize throughput, new pixels are stored where the now obsolete pixels were located in memory without reordering them. Since the individual tiles overlap by exactly $2/3$ the tile is subdivided into squares of $16 \times 16$ pixels (visualized in \figref{fig:checkerboard_pixelPerm}). The rows and columns of this $3 \times 3$ grid undergo a cyclic permutation that results in 9 different fragmentation states. The \emph{filter cluster} and the \emph{merger} keep track of the fragmentation state and transform the addresses accordingly. This approach increases the complexity of the address calculation but minimizes pipeline stalls in the \emph{filter cluster} and allows 2/3 of the tiled pixels to remain in the SRAMs when stepping to the next tile.
%
%
\section{Implementation \& Results}
\label{sec:results}
\begin{figure}[t]
    \centering
    \includegraphics[width=0.36\textwidth]{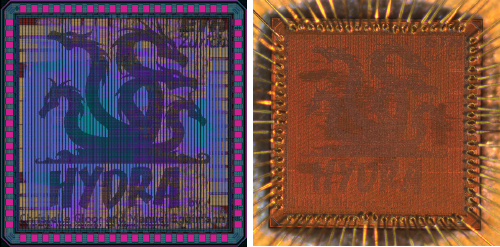}
    \vspace{-2mm}       
    \caption{CAD rendering and microphotograph of the \textsc{Hydra} ASIC.}
    \label{fig:hydra_images}
    \vspace{-4mm}  
\end{figure}

A chip named \textsc{Hydra} (depicted in \figref{fig:hydra_images}) implements the proposed architecture and was fabricated in \SI{65}{\nano\metre} CMOS technology. The design has been synthesised with Synopsys DC 2016.03 and P\&R has been performed with Cadence Innovus 16.1. The total design complexity is 1.3\,MGE of which 43\% is occupied by the 52 SRAMs. \textsc{Hydra} supports at-runtime configuration of the filter parameters and arbitrary video resolutions (with real-time performance up to 720p at 1.2\,V). 
It also features a high FP24 compute density of \SI{6.7}{\giga FLOPS/\milli\metre\squared}. When scaled to \SI{16}{\nano\metre}, this would amount to \SI{89}{\giga FLOPS/\milli\metre\squared}, which is around 12$\times$ higher than in modern mobile GPUs manufactured in \SI{16}{\nano\metre}. For instance, the NVidia Tegra X2 provides a theoretical peak throughput of \SI{750}{\giga FLOPS} \cite{Skende2016}, and with an assumed silicon area of \SI{100}{\milli\metre\squared} for the GPU subsystem, this results in only \SI{7.5}{\giga FLOPS/\milli\metre\squared}. In terms of external memory bandwidth, \textsc{Hydra} requires \SI{950}{\mega\byte/\second}. This amounts to only 7.4\,\% of the total bandwidth provided by a LPDDR4-1600 \SI{64}{\bit} memory channel, which makes our design an ideal candidate for inclusion within a specialized domain accelerator in a SoC. 

\tabref{tbl:key-figures} compares the key figures of \textsc{Hydra} with two related accelerators \cite{Tseng2011,Kao2014}. Note, that these designs implement simpler EA filters (BF/GF variants) with fixed
point arithmetic since they have been developed to process dense, 8\,bit standard dynamic range (SDR) data occurring in applications like flash denoising \cite{Kao2014}. Our design is the only one that reports measured data, and it has been designed to support much more challenging HDR images and sparse optical flow data, which requires accurate arithmetic with high dynamic range (\secref{sec:precision}). The PF provides better filtering quality than the GF and BF since it does not suffer from halo artifacts \cite{Aydin2014}.
%
\begin{table}[t]
\begin{threeparttable}
\small
\newcommand{\LY}{\checkmark}
\newcommand{\LN}{--}
\newcommand{\NFB}{$\rightleftarrows$}
\newcommand{\NF}{$\rightarrow$}
\newcommand{\NB}{$\leftarrow$}
\newcommand{\NN}{--}
\renewcommand{\arraystretch}{0.7}
\caption{Key figures of \textsc{Hydra} and comparison with related designs optimized for applications that only work on 8\,bit SDR data.\vspace{-2mm}}
\label{tbl:key-figures}
\centering
\begin{tabularx}{\linewidth}{@{}p{0.4cm}@{}|p{3.3cm}@{}|@{}x{1.68cm}@{}x{1.68cm}@{}x{1.65cm}@{}}
\toprule
\multicolumn{2}{c|}{\textbf{Properties\hspace{4mm}\textbackslash{}\hspace{4mm}Design}}& \cite{Tseng2011}          & \cite{Kao2014}            & \textsc{Hydra}          \\
\midrule
\parbox[t]{2mm}{\multirow{5}{*}{\rotatebox[origin=c]{90}{\textbf{\hspace{6mm}Algo.}}}}
& {Filter Type}                     & Joint BF                  & GF                        & TPF                     \\
& {Window Size [px]}                & 31$\times$31              & 31$\times$31              & 48$\times$48            \\
& {Arithmetic}                      & FIXP                      & FIXP                      & FP24                    \\
\midrule 
\parbox[t]{2mm}{\multirow{5}{*}{\rotatebox[origin=c]{90}{\textbf{\hspace{6mm}Appl.}}}}
& SDR Data                    & \LY                 & \LY                 & \LY               \\
& HDR Data                    & \LN                 & \LN                 & \LY               \\
& Sparse Coordinate Data      & \LN                 & \LN                 & \LY               \\
\midrule
\parbox[t]{2mm}{\multirow{5}{*}{\rotatebox[origin=c]{90}{\textbf{Resources}}}}
& {Results from}                    & Gate-Level                & Post-Layout               & Measured                \\
& {Technology [nm]}                 & 90                        & 90                        & 65                      \\
& {Logic [kGE]}                     & 276                       & 93                        & 762                     \\
& {SRAM [kB]}                       & 23                        & 3.2                       & 47.3                    \\
& {Total Complexity [kGE]}          & -                         & -                         & 1'328                   \\
\midrule 
\parbox[t]{2mm}{\multirow{5}{*}{\rotatebox[origin=c]{90}{\textbf{Perform.}}}}
& {Resolution}                      & 1080p                     & 1080p                     & 720p                    \\
& {Frequency  [MHz]}                & 100                       & 100                       & 259                     \\
& {Throughput [fps]}                & 30                        & 30                        & 24.8                    \\
& {Core Power [mW/MHz]}             & -                         & 0.23                      & 1.72                    \\
& {Bandwidth [MB/frame]}            & 16.6                      & 32.8                      & 38                      \\
\bottomrule
\end{tabularx}
\begin{tablenotes}
\end{tablenotes}
\end{threeparttable}
\vspace{-3mm}
\end{table}

%
\section{Conclusions}
\label{sec:conclusion}
We present \textsc{Hydra}, a compact and efficient accelerator for high-quality, permeability-based EA filtering in real-time. The accelerator employs a novel, tiled variant of the PF that significantly reduces the required on-chip memory and off-chip bandwidth compared to a non-tiled PF implementation. \textsc{Hydra} is parameterized to deliver a throughput of 25\,fps for monochromatic 720p video and provides significantly higher compute density than recent mobile GPUs. Our design is scalable to higher resolutions by increasing the amount of parallel FUs, and by employing higher-order pipeline interleaving to increase the operating frequency. Integrated into a SoC, the presented accelerator could act as a highly accurate and area-efficient co-processor for video processing applications. 
%
\newpage
\bibliographystyle{IEEEtran}
\bibliography{bibliography}

\begin{thebibliography}{10}
\providecommand{\url}[1]{#1}
\csname url@samestyle\endcsname
\providecommand{\newblock}{\relax}
\providecommand{\bibinfo}[2]{#2}
\providecommand{\BIBentrySTDinterwordspacing}{\spaceskip=0pt\relax}
\providecommand{\BIBentryALTinterwordstretchfactor}{4}
\providecommand{\BIBentryALTinterwordspacing}{\spaceskip=\fontdimen2\font plus
\BIBentryALTinterwordstretchfactor\fontdimen3\font minus
  \fontdimen4\font\relax}
\providecommand{\BIBforeignlanguage}[2]{{%
\expandafter\ifx\csname l@#1\endcsname\relax
\typeout{** WARNING: IEEEtran.bst: No hyphenation pattern has been}%
\typeout{** loaded for the language `#1'. Using the pattern for}%
\typeout{** the default language instead.}%
\else
\language=\csname l@#1\endcsname
\fi
#2}}
\providecommand{\BIBdecl}{\relax}
\BIBdecl

\bibitem{Tomasi1998}
C.~Tomasi and R.~Manduchi, ``Bilateral filtering for gray and color images,''
  in \emph{ICCV}, Jan 1998, pp. 839--846.

\bibitem{Perona1990}
P.~Perona and J.~Malik, ``Scale-space and edge detection using anisotropic
  diffusion,'' \emph{IEEE TPAMI}, vol.~12, no.~7, Jul 1990.

\bibitem{Milanfar2013}
P.~Milanfar, ``A tour of modern image filtering: New insights and methods, both
  practical and theoretical,'' \emph{IEEE SPM}, Jan 2013.

\bibitem{Farbman2008}
Z.~Farbman, R.~Fattal, D.~Lischinski \emph{et~al.}, ``Edge-preserving
  decompositions for multi-scale tone and detail manipulation,'' \emph{ACM
  TOG}, vol.~27, no.~3, pp. 67:1--67:10, Aug. 2008.

\bibitem{Fattal2009}
R.~Fattal, ``Edge-avoiding wavelets and their applications,'' \emph{ACM TOG},
  vol.~28, no.~3, pp. 1--10, 2009.

\bibitem{He2013}
K.~He, J.~Sun, and X.~Tang, ``Guided image filtering,'' \emph{IEEE TPAMI},
  vol.~35, no.~6, pp. 1397--1409, June 2013.

\bibitem{Gastal2011DT}
E.~S.~L. Gastal and M.~M. Oliveira, ``Domain transform for edge-aware image and
  video processing,'' \emph{ACM TOG}, vol.~30, no.~4, 2011.

\bibitem{Gastal2015}
E.~S. Gastal and M.~M. Oliveira, ``{High-Order Recursive Filtering of
  Non-Uniformly Sampled Signals for Image and Video Processing},'' in
  \emph{Computer Graphics Forum}, vol.~34, no.~2, 2015, pp. 81--93.

\bibitem{Aubry2014}
M.~Aubry, S.~Paris, S.~W. Hasinoff \emph{et~al.}, ``{Fast Local Laplacian
  Filters: Theory and Applications},'' \emph{ACM TOG}, vol.~33, no.~5, 2014.

\bibitem{Paris2011}
S.~Paris, S.~W. Hasinoff, and J.~Kautz, ``{Local Laplacian Filters: Edge-Aware
  Image Processing With a Laplacian Pyramid},'' \emph{ACM TOG}, vol.~30, no.~4,
  p.~68, 2011.

\bibitem{cigla2013}
C.~Cigla and A.~A. Alatan, ``{Information Permeability for Stereo Matching},''
  \emph{Signal Processing: Image Communication}, 2013.

\bibitem{Aydin2014}
T.~O. Aydin, N.~Stefanoski, S.~Croci \emph{et~al.}, ``{Temporally Coherent
  Local Tone Mapping of HDR Video},'' \emph{ACM TOG}, 2014.

\bibitem{Schaffner2017}
M.~Schaffner, F.~Scheidegger, L.~Cavigelli \emph{et~al.}, ``{Towards Edge-Aware
  Spatio-Temporal Filtering in Real-Time},'' \emph{IEEE TIP}, vol.~PP, 2017.

\bibitem{Lang2012}
M.~Lang, O.~Wang, T.~Ayd{\i}n \emph{et~al.}, ``{Practical Temporal Consistency
  for Image-Based Graphics Applications},'' \emph{ACM TOG}, vol.~31, no.~4,
  2012.

\bibitem{Porikli2008}
F.~Porikli, ``Constant time o(1) bilateral filtering,'' in \emph{IEEE CVPR
  2008}, June 2008, pp. 1--8.

\bibitem{Tseng2011}
Y.~C. Tseng, P.~H. Hsu, and T.~S. Chang, ``{A 124 Mpixels/s VLSI Design for
  Histogram-Based Joint Bilateral Filtering},'' \emph{IEEE TIP}, vol.~20,
  no.~11, pp. 3231--3241, Nov 2011.

\bibitem{Kao2014}
C.~C. Kao, J.~H. Lai, and S.~Y. Chien, ``{VLSI Architecture Design of Guided
  Filter for 30 Frames/s Full-HD Video},'' \emph{IEEE TCSVT}, vol.~24, no.~3,
  pp. 513--524, March 2014.

\bibitem{Kaeslin2014}
H.~Kaeslin, ``{Top-Down Digital VLSI Design, from VLSI Architectures to
  Gate-Level Circuits and FPGAs},'' \emph{Morgan Kaufmann}, 2014.

\bibitem{Skende2016}
A.~Skende, ``Introducing parker next-generation tegra system-on-chip,'' in
  \emph{2016 IEEE Hot Chips 28 Symposium (HCS)}, Aug 2016, pp. 1--17.

\end{thebibliography}

\end{document}